\documentclass[aps,pre,twocolumn,superscriptaddress,showpacs,floatfix]{revtex4}
\usepackage{amssymb}
\usepackage{epsfig}
\usepackage{amsmath}
\usepackage{times}
\usepackage{color}
\usepackage[usenames,dvipsnames,svgnames,table]{xcolor}

\begin{document}

\title{Entrainment of the suprachiasmatic nucleus network by a light-dark cycle}

\author{Jinshan Xu}
\affiliation{Department of Physics, East China Normal University,
Shanghai 200062, China}
\affiliation{Laboratoire de Physique, ENS
de Lyon and CNRS, 46 All$\acute{e}$e d'Italie, 69007, Lyon,
France}

\author{Changgui Gu}
\affiliation{Department of
Physics, East China Normal University, Shanghai 200062, China}

\author{Alain Pumir}
\affiliation{Laboratoire de Physique, ENS de Lyon and CNRS, 46
All$\acute{e}$e d'Italie, 69007, Lyon, France}

\author{Nicolas Garnier}
\email{nicolas.garnier@ens-lyon.fr} \affiliation{Laboratoire de Physique, ENS de Lyon and CNRS, 46
All$\acute{e}$e d'Italie, 69007, Lyon, France}

\author{Zonghua Liu}
\email{zhliu@phy.ecnu.edu.cn} \affiliation{Department of Physics,
East China Normal University, Shanghai 200062, China}

\date{\today}

\begin{abstract}
The synchronization of biological activity with the
alternation of day and night (circadian rhythm) is performed in the brain
by a group of neurons, constituting
the suprachiasmatic nucleus (SCN). The SCN is divided into two subgroups of
oscillating cells: the ventro-lateral (VL) neurons, which are exposed to
light (photic signal) and the dorso-medial (DM) neurons which are coupled to the VL cells.
When the coupling between these neurons is strong enough, the system synchronizes with the
photic period. Upon increasing the cell coupling, the entrainment of the DM cells has been
recently shown to occur via a very sharp (jumping) transition when the period
of the photic input is larger than the intrinsic period of the cells.
Here, we characterize this transition with a simple realistic model. We show
that two bifurcations possibly lead to the disappearance of the endogenous
mode. Using a mean field model, we show that the jumping transition
results from a supercritical Hopf-like bifurcation. This finding implies that both the
period and strength of the stimulating photic signal, and the relative fraction
of cells in the VL and DM compartments are crucial in determining the
synchronization of the system.

\end{abstract}



\maketitle

\section{Introduction}
The circadian rhythm in the brain results from the activity of neurons, which
are spontaneously oscillating with an endogenous period close to the $24$ hour
cycle, and which, under the influence of light (photic signal), synchronize
with the alternation of day and night.
%
In mammals, the primary circadian clock is the suprachiasmatic nucleus (SCN), located in the hypothalamus and receiving information about illumination through the eyes.
It is composed of a large network of $\sim 2 \times 10^4$ coupled neurons. This
assembly of cells can be divided into two subgroups: the ventrolateral (VL)
and the dorsomedial (DM) subgroups. The VL neurons 
are exposed to the photic
input from the retina and entrain the DM neurons.
While the two subgroups of neurons are functionally different, a coherent,
periodic output results from their coupling~\cite{Gonze:2005,Bernard:2007, Locke:2008,Daido:2001,Li:2006, Leloup:1999,Ruoff:1996,Ruoff:2001,Welsh:2010,Gu:2009,Gu:2011}.

In the absence of the daily light-darkness cycle, the free-running period varies
from species to species in the range $20-28$h \cite{Daido:2001,Gonze:2005,Bernard:2007,Leloup:1999,Ruoff:1996,Ruoff:2001}.
This implies that the proper response of the system to the $24$h period
results from a dynamic process of synchronization.
In this respect, it has been noticed that when exposed for several weeks to a
constant light, the SCN of rodents (hamster) exhibits a phase-splitting
behavior, with two
sets of neurons oscillating out of phase~\cite{Pitten:1976,Pitten:1993,Pavlidis:1978,Iglesia:2000,Ohta:2005,Yan:2005}.
It has been recently shown that both the coupling strength and its distribution
can influence the diversity of free-running period and the phase-splitting
\cite{Gu:2009,Gu:2011}.
Last, the desynchronization of the circadian oscillations
between VL and DM subgroups has been observed when the external
light-dark cycle has a period very different from the $24$h
circadian period
\cite{Albus:2005,Iglesia:2004,Nagano:2003,Nakamura:2005,Gu:2012}.
When the period of light-dark cycle is smaller than $24$h such as
22h (11h of light alternating with 11h of darkness), the VL subgroup is
entrained by the light and oscillates with a period equal to the
external cycle (22h), whereas the DM subgroup is not entrained and oscillates
with its free-running period around 24h, as observed experimentally on
rats \cite{Iglesia:2004,Campuzano:1998}.
In the opposite case where the period of light-dark cycle is longer than 24h, such as 26h (13h of light, alternating with 13h of darkness), numerical simulations \cite{Gu:2012} predict that
the VL subgroup is entrained by the light but the DM subgroup has a period
smaller than 24h. By gradually increasing numerically the number of neurons
in the VL subgroup, the period of the DM subgroup is observed to decrease.
Theoretically, the entrainment phenomenon has been analyzed in terms of frequency locking in the case of a homogeneous VL population~\cite{Granada:2011}. In the case of an heterogeneous population with both DM and VL neurons, one of the intriguing observations of \cite{Gu:2012} is the existence
of a threshold for the ratio between the number of neurons in the VL and DM.
When the ratio reaches a critical value, the period of DM
subgroup jumps to the external light-dark cycle.  
We focus here on this jumping transition phenomenon, which
provides new insight on the entrainment of photic input in SCN,
i.e., the appearance of rhythm.

To investigate the mechanism of the jumping transition in the SCN, we use a
model with a mean-field coupling 
and characterize both the period and amplitude in each subgroup. We find
that as the fraction of VL neurons is increased, the period of the DM
subgroup decreases and so does its amplitude. Based on this finding, we study
a single oscillator with both constant light and monochromatic light.
We find that this model shows the very same behavior as the DM subgroup,
including the disappearance of amplitude via a Hopf-like bifurcation,
which explains the jumping transition. The main implication concerning
the generation and synchronization of rhythm in the SCN network is
that both the period and strength of the stimulating photic signal,
and the relative fraction of cells in the VL and DM compartments
are crucial in determining the synchronization of the system.

\section{A jumping transition in Goodwin model}

A typical model to simulate circadian rhythms in SCN cells is the Goodwin
oscillator with three variables, describing a negative
transcription-translation feedback loop \cite{Goodwin:1965}.
Several elaborations of this models have been proposed
\cite{Gonze:2005,Bernard:2007,Leloup:1999,Ruoff:1996,Ruoff:2001}.
For example, \cite{Gonze:2005,Bernard:2007} considered a global coupling
strength depending on the concentration of neurotransmitter into
the Goodwin oscillator and made them be influenced under a
mean-field.
Locke {\it et al.} modified the system to coupled damped oscillators
\cite{Locke:2008}. We consider here the mean-field Goodwin oscillator
presented by Gonze {\it et al.} \cite{Gonze:2005}, which can be
represented as follows
\begin{eqnarray}\label{eq:Gonze}
\dot{x}_i&=&\frac{\alpha_1}{1+(z_i/k_1)^4}-\frac{\alpha_2 x_i}{k_2+x_i}
+\frac{\alpha_c gF}{k_c+gF}+L_i, \nonumber \\
\dot{y}_i&=&k_3x_i-\frac{\alpha_4y_i}{k_4+y_i},\nonumber \\
\dot{z}_i&=&k_5y_i-\frac{\alpha_6z_i}{k_6+z_i},\nonumber \\
\dot{V}_i&=&k_7x_i-\frac{\alpha_8V_i}{k_8+V_i}, \quad i=1,2,\cdots,N  \\
F&=&\frac{1}{N}\sum_{i=1}^NV_i, \nonumber
\end{eqnarray}
The variables $x_i,$ $y_i$, and  $z_i$  are the concentrations of clock gene mRNA,  clock protein and inhibitor of protein expression respectively~\cite{Gonze:2005}. $V$ is the concentration of neuropeptide induced by the activation of the clock gene  and can synchronize clock cells. The three variable model obtained with $x_i, y_i$ and $z_i$ constitute a negative feedback loop in the clock cell $i$.
$g$ measures the strength of the mean field coupling $F$, and $L$ denotes the external light input. 
 Following Ref. \cite{Locke:2008} we take other
parameters as $\alpha_1=0.7$nM/h, $k_1=1.0$nM, n=4.0,
$\alpha_2=0.35$nM/h, $k_2=1.0$nM,  $k_3=0.7$/h,
$\alpha_4=0.35$nM/h, $k_4=1.0$nM,  $k_5=0.7$/h,
$\alpha_6=0.35$nM/h, $k_6=1.0$/h, $k_7=0.35$/h,
$\alpha_8=1.0$nM/h, $k_8=1.0$nM,  $\alpha_c=0.4$nM/h,
$k_c=1.0$nM. We renormalize the time by a factor $1.26$ to have a free-running period of 24h, and we use $g=0.5$.
 
{Our model of the SCN network is composed of $N$ oscillators,}
all obeying Eq.~(\ref{eq:Gonze}) and coupled together via the mean field $F$. 
{The observation of~\cite{Gu:2012} that the variability of the coupling constant $g$ in the model does not affect qualitatively the nature of the transition strongly suggests that such an approximation is sensible.}
The assembly of neurons is divided into two subgroups.
The oscillators in the VL subgroup receive photic input $L_i = L(t)$,
while oscillators in the DM subgroup do not receive any light (hence $L_i=0$).
The fraction of VL neurons in the SCN network is noted $p$, so the system
consists of $pN$ VL neurons, and $(1-p)N$ DM neurons. We study the entrainment
of the system by an alternation of day and light, with a period of 26h,
with a
photic input $L(t)$
chosen to be on for 13h: $L(t) = K$ during the day period, and off for
13h : $L(t) = 0$ during the period of darkness.
%
\begin{figure}
\epsfig{figure=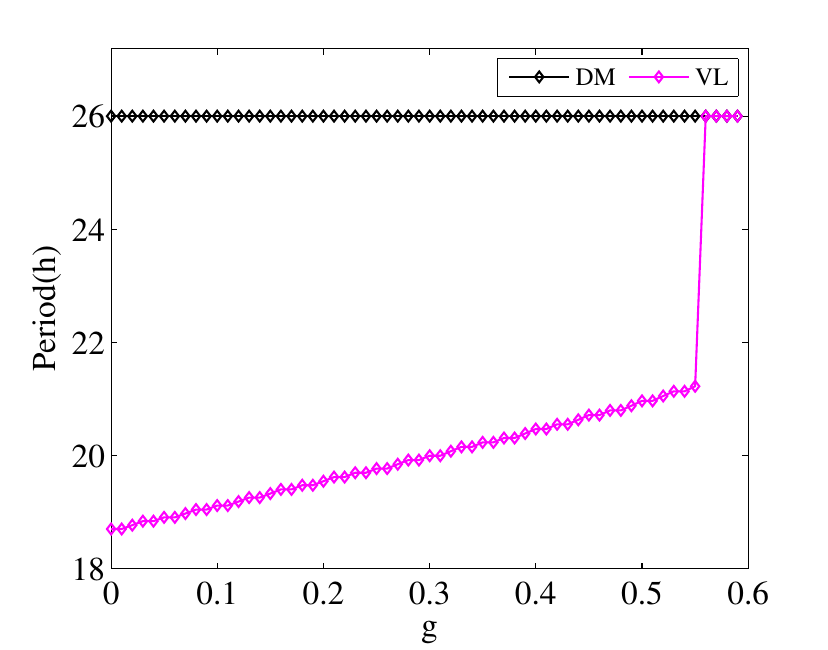,width=1.0\linewidth}
\caption{(color online). Evolution of the period of the VL ({\color{black} black $\diamond$}) and DM ({\color{magenta} magenta $\circ$}) subgroups in a system with a fraction $p$ of VL neurons. The VL neurons are exposed to a 26h-cycle light of intensity $K=0.02$. System is evolved from random initial conditions.
Symbols represent a system of $N=500$ neurons, and continuous lines represent a 2-neuron system (labeled $N_\infty$). }
\label{fig:VL_DM_periods_vs_p}
\end{figure}
We checked the behaviors of $\{x_i\}$ in Eq.~(\ref{eq:Gonze}) and found that
the DM oscillators are completely synchronized with one another.  We observed that this was the case for any value of $p$.
We define the period of a subgroup (VL or DM) as the period of the average $V$ variable in the subgroup. 
{After a long transient, all the oscillators in each sub-population become synchronized. Their frequencies were determined numerically by computing the Fourier spectrum.}
%
Fig.~\ref{fig:VL_DM_periods_vs_p} shows how the period of the
subgroups evolves with $p$, which is an important factor for the entrainment of the period of photic input~ \cite{Gu:2012}.
As $p$ is increased from 0, Fig. \ref{fig:VL_DM_periods_vs_p} shows
that the period of VL is quickly frequency-locked to the external period (26h) 
in a 1:1 relation for a small value of $p$ ~\cite{Granada:2011}, whereas the period of DM decreases with increasing
$p$ until a critical value $p_c=0.41$ is reached. At the value $p = p_c$, we observe
a transition, characterized by a sharp discontinuity in the period
of the DM neurons, which jumps from $ \approx 20.8$h to the external light
period 26h. At values of $p \ge p_c$, the DM neurons are entrained at
the external period, 26h.
{We have observed the jumping transition at different values of the coupling constant g, and also by varying g at a fixed value of p.}
What is the mechanism describing this jumping transition? To gain some
insight, we measured the amplitude of the oscillations of DM oscillators.

Because of the mean field structure of the coupling between neurons, together
with the observation that neurons from a given subgroup are perfectly synchronized in
this subgroup for any value of $p$~\cite{Gu:2012}, all oscillators from
a subgroup can be treated as a single oscillator. It is therefore sufficient
to study a two-neurons system, composed of one VL neuron receiving external
light $L(t) \ne 0$ and one DM neuron insensitive to light, both being coupled
by the mean field $F=pV_{\rm VL}+(1-p)V_{\rm DM}$.
Although this approach can potentially lead to incorrect results close to the
transition, due to the divergence of the relaxation time, we explicitly checked that this is not the case for reasonable integration times (500h).
Fig.1 shows that the jumping transition occurs for the same value of $p$ in the
$2$-neuron system as well as in a large $N$ system.
%

\begin{figure}
\epsfig{figure=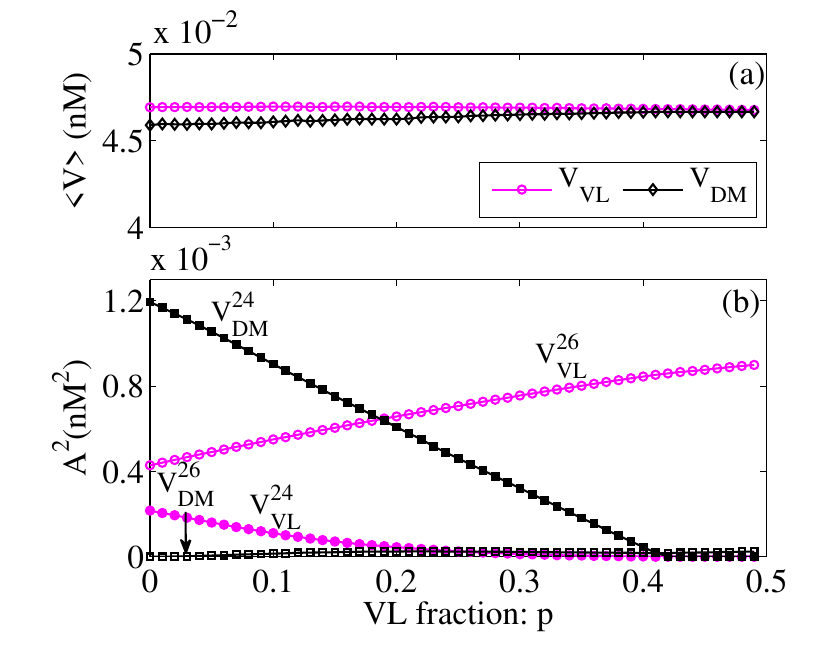,width=1.0\linewidth}
\caption{(color online) Evolution with $p$ of the $V$ variable in a 2-neurons system. (a): Stationary component $\langle V\rangle$ for the VL ({\color{magenta} magenta $\circ$}) and DM populations (black $\diamond$). (b): Square amplitude of the two main oscillating modes of VL and DM populations. The forcing mode has a period of 26h, while the free-running mode has a period ranging from 24h down to 20.8h, according to Fig.\ref{fig:VL_DM_periods_vs_p}}.
\label{fig:mean-field}
\end{figure}
Considering that the mean field term $F$ is the only coupling between VL and
DM neurons, and willing to study the effect of the fraction $p$ of cells in the two subgroups,
we study the dependence on $p$ of $V_{\rm VL}$ and $V_{\rm DM}$.
To this end, we decomposed for each subgroup the variable $V$ into three parts:
the stationary value $\langle V\rangle$, the oscillating components
$V^{26}$ corresponding to a period of exactly 26h, {\it i.e.}, the forcing period,
and $V^{24}$ corresponding to the branch that starts at 24h (free-running mode), and diminishes down to 20.8h at the transition for the DM population. The amplitudes of $V^{26}$
and $V^{24}$ are measured by integrating the power spectrum density under the
corresponding peaks. We observe in Fig.~\ref{fig:mean-field} that the constant
part of the variable $V$ does not depend much on $p$ and that it is
approximately the same, up to at most a $2\%$ variation, for both the VL and
the DM neuron. Therefore we can consider $\langle F \rangle$ as a constant.
The amplitude $V_{\rm VL}^{26}$ of the mode at period 26h increases
for  the VL neurons, due to the increase of the ratio of VL neurons in the system. For DM neurons, $V_{\rm DM}^{26}$ also increases with $p$ but remains very small.
In contrast, the
$V_{\rm VL}^{24}$ mode in the VL neuron has a decreasing amplitude with
$p$, while the amplitude of the free-running mode of the DM neuron starts,
at $p = 0$, at a much larger value, and vanishes close to $p = p_c$
like $\propto (p-p_c)^{1/2}$ (squared amplitude is plotted in
Fig.~\ref{fig:mean-field}(b)) which strongly suggests a supercritical Hopf
bifurcation {for the $V_{\rm VL}^{24}$ mode and therefore a torus bifurcation or Neimark-Sacker bifurcation~\cite{Sacker:08} for the complete system}. From the definition of $F$ and the values of $V_{\rm VL}$ and $V_{\rm DM}$ in Fig.~\ref{fig:mean-field}(a) we note that the mean value of $F$ 
is always positive and large. In fact, the amplitude of the oscillating part
is small, so $F$ remains strictly positive at all times. This is to be
contrasted with the photic input $L$ which is zero for half a period.

The main effect of the mean field is to transfer the information about the
external forcing from the VL population to the DM population. As $p$ is
increased, the mean field contains more information about the VL population and
the transfer to the DM population is more efficient. This leads to the
disappearance of the free-running mode in the DM subgroup via a Neimark-Sacker or Hopf-like bifurcation.
In the next section, we explore the transition by studying the effect of the external forcing on a single Goodwin oscillator.

\section{Single neuron analysis}

To understand the entrainment of a given subgroup by the photic signal, it is
useful to consider first the simplified case of a single neuron,
subject to a periodic photic forcing $L(t) = L_0 + L_1 \sin(\omega t)$,
with  $2\pi/\omega=26$ hours.
In general, the behavior of neurons of either category (VL or DM) is determined
by a balance between the coupling with the mean field $F=V$,
and the external forcing. For this reason, we investigate the dependence of
the dynamics on the parameters $g$, the strength of the coupling, $L_0$, the amplitude of the
constant forcing and $L_1$, the amplitude of a periodic monochromatic forcing
with a period $26h$.
We will then use our knowledge of this single isolated neuron system to describe the behavior of VL and DM populations in a network : external light, together with a contribution
from the mean field, will be assimilated to $L(t)$.

\subsection{VL population}

\begin{figure}
\epsfig{figure=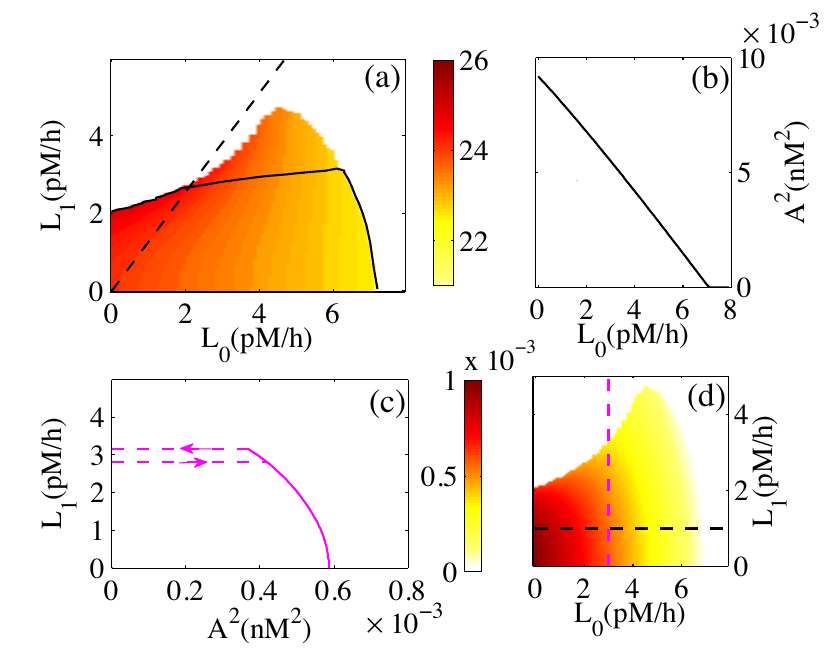,width=1.0\linewidth}
\caption{(color online) Period and square amplitude of the oscillation of a single neuron model. (a): Period as a function of $(L_0, L_1)$ for fixed $g$=0.5. White color corresponds here to the forcing period 26h (the neuron is entrained by the external light). Dotted line represents the situation where the system is subject to a periodic on-off light. {The solid line delimits two subregions for the endogeneous mode : below the solid line, the endogeneous oscillation is the only solution whereas above the solid line there is bistability and both the endogeneous solution and the entrained solution can be observed depending on initial conditions.}
(b): Supercritical Hopf bifurcation occurs when reducing $L_0$ for fixed $L_1$=1pM/h. (c): 
{Subcritical is revealed by an hysteresis and bistability in the squared amplitude}
when increasing $L_1$ for fixed $L_0${=3pM/h}. (d) : squared amplitude of the free-running mode in the $(L_0,L_1)$ plane for $g$=0.5. Dotted lines in (d) give the location of the cuts plotted in (b) and (c). 
}
\label{fig:one_neuron_fixed_g}
\end{figure}

We first study a single VL neuron at a fixed value
of the coupling, $g=0.5$ (Fig.~\ref{fig:one_neuron_fixed_g}). Over a wide
range of values of $L_0$ and $L_1$, the system is observed to oscillate.
Fig.~\ref{fig:one_neuron_fixed_g}(a) shows the dependence of the period of
oscillations on $L_0$ and $L_1$. 
{For either large $L_0$ or large $L_1$,}
the VL neuron is 
{entrained by }
the external light, while it keeps its free-running
period for smaller values of $L_0$ and $L_1$.
%

Fig.~\ref{fig:one_neuron_fixed_g}d shows the dependence of the squared
amplitude of the oscillation. We observe that an increase of the constant
part of the forcing (parameter $L_0$), for fixed periodic monochromatic forcing (constant $L_1$)
leads to the vanishing of the amplitude of the free-running mode, while its
frequency decreases (from 24h down to about 21h). More precisely, the squared amplitude vanishes linearly with the control parameter $L_0$ (see Fig.~\ref{fig:one_neuron_fixed_g}b) This demonstrates that 
{the transition from a non-oscillatory to an oscillatory state,}
when $L_0$ decreases, happens through a supercritical Hopf bifurcation.
On the contrary, an increase of the amplitude of the forcing at the period
26h (parameter $L_1$) at fixed value of $L_0$ leads to
the abrupt disappearance of the free-running mode (see Fig.~\ref{fig:one_neuron_fixed_g}c). The fact that the
oscillating solution ceases to exist, while the amplitude of the oscillation
is non-zero, suggests a subcritical bifurcation.

This can be used to describe qualitatively the behavior of the VL population
in the complete SCN model, where external forcing is chosen as an alternation
of darkness ($L=0$) for 13h, followed by light ($L=K$ for 13h), which can be decomposed in Fourier series. The constant term is equal to
$L_0 = K/2$ and the first harmonic, with period 26h, has an amplitude
$L_1 = 4 L_0 /\pi$ (dotted line in Fig.~\ref{fig:one_neuron_fixed_g}a). The higher harmonics have a frequency which is too high to trigger a significant response. In more technical term, the frequencies
corresponding to harmonics of frequency $n \times 2 \pi/T$, with $T =$ 26h,
are very far from the resonance tongue of the Goodwin
oscillator~\cite{Arnold:1996,Granada:2011} as soon as $n>1$.

Qualitatively, we expect VL neurons in a network to behave as a single neuron
receiving a photic signal $L(t)$, in addition to a contribution
from the mean field. In the simulation above, it was found that
$\langle F \rangle$ is independent of $p$ (see Fig.~\ref{fig:mean-field}a). As a result,
as $p$ is increased, the VL subgroup behaves like a one-neuron system with
constant $L_0$. On the contrary, $V_{\rm DM}^{26}$ is negligible compared to
$V_{\rm VL}^{26}$ which increases with the fraction $p$
(see Fig.~\ref{fig:mean-field}b), so the mean field $F$ will contain
a mode at a period of $26$ h, which will grow like
$p V_{\rm VL}^{26}$ when $p$ is increased. This suggests that the VL subgroup behaves like a one-neuron system with increasing $L_1$ when $p$ increases. According to Fig.~\ref{fig:one_neuron_fixed_g}d,
for fixed $L_0$ and increasing $L_1$, we expect the
bifurcation to the 26h state to be subcritical.
 This corresponds to the
transition of the VL population observed at $p=0$ in
Fig.~\ref{fig:VL_DM_periods_vs_p}, which occurs for larger values of $p$ if
the light intensity $K$ is smaller~\cite{Gu:2012}.

\subsection{DM population}
We now turn to the main object of this article, the DM subgroup, and show
quantitatively that the jumping transition to the external forcing frequency is a supercritical Hopf bifurcation.
%
\begin{figure}
\epsfig{figure=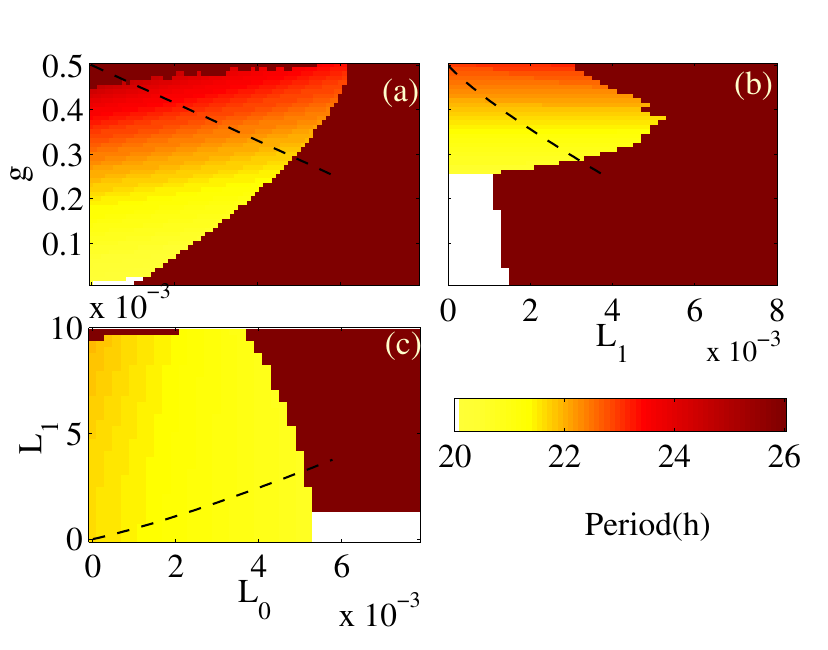,width=1.0\linewidth}
\caption{(color online) Period of a single neuron as a function of : (a) $(L_0,g)$ for fixed $L_1$=0.0032, (b): $(L_1,g)$ for fixed $L_0$=0.0048, and (c) $(L_0,L_1)$ for fixed g=0.295. Dotted lines are projections of the trajectory of the system for increasing $p$. White regions correspond to a non-oscillating neuron.}
\label{fig:one_neuron}
\label{fig:Hopf_bifurcation}
\end{figure}
The DM oscillators are not directly forced by the 26h-periodic light, but indirectly
{\em via} the mean field $F$. From the simulations reported
previously, see Fig.~\ref{fig:mean-field}, we find that
$gF \simeq 3\times10^{-2}$ nM is always much smaller than $k_c=1.0$nM, so
the coupling term can be linearized, and rewritten by introducing an
effective external light:
\begin{equation}
\frac{\alpha_c gF}{k_c+gF} \simeq \frac{\alpha_c g}{k_c} F = \frac{\alpha_cg'}{k_c} V_{DM} + L_0 + L_1 \sin(\omega t)\,,
\end{equation}
with an effective coupling $g' = (1-p)g$, and an external light $L(t)$
proportional to $p V_{VL}(t)$,
with temporal average $L_0 = \alpha_c g p \langle V_{VL} \rangle/k_c$ and amplitude $L_1=\alpha_c g p V_{VL}^{26}/k_c$.
Fig.\ref{fig:one_neuron} presents the dynamical behavior of a single neuron
as the three parameters are varied. In this phase space, the trajectory $(g'(p), L_0(p), L_1(p))$ of the complete
SCN system obtained by increasing $p$ is deduced from the analytical
expressions of the effective parameters, by using $\langle V_{VL} \rangle$
and $V_{VL}^{26}$ from Fig.\ref{fig:mean-field}.
The projections of this trajectory are plotted as a dotted lines in
Fig.~\ref{fig:Hopf_bifurcation}.
We observe that the transition of the one-neuron system is a supercritical
Hopf bifurcation, exactly as the one in Fig.~\ref{fig:one_neuron_fixed_g}b for
fixed $g$. This bifurcation occurs for an effective coupling $g'=0.295$, from which we deduce $p_c=0.41$, in perfect agreement with the values observed in the complete SCN
network (Fig.~\ref{fig:VL_DM_periods_vs_p}). The period of the free-running
mode at the transition (20.8h) is also the same in this 1-neuron analysis as in the complete SCN network.


\section{Discussions and conclusions}

We have described the entrainment of both the VL neurons and the the DM neurons by a simple model of a single neuron forced by an external light $L(t)$. This forcing is a key element in the description of the Goodwin oscillator. The mean field coupling between subpopulations blurs the distinction between VL and DM oscillators which allows us to treat the DM neurons as VL neurons under specific light, and show that the DM neurons synchronize with the external light, although they are not directly coupled to it.

We documented that the sharp transition from the endogenous period of the DM population to the external period of the forcing corresponds to a supercritical Hopf-like bifurcation for the endogenous mode. For a ratio $p$ larger than the critical value $p_c$, there are no more oscillations at a period different from the forcing period (26h). Nevertheless, both populations of neurons oscillate and sustain the external period. The DM population is then slaved to the mean field $F$ which only contains the 26h mode, while the VL population follows the external forcing. So, in terms of the ratio $p$, we can predict that the amplitude of the DM oscillations for $p>p_c$ do not vanish but is roughly proportional to the mean field $F$, {\em i.e.}, to $p$ itself, as observed in Fig.5d of \cite{Gu:2012}.

In conclusion, we have isolated two possible transitions to explain the entrainment of the VL and DM populations by an external light under mean field coupling. The jumping transition of the DM subgroup depicted in Fig.~\ref{fig:VL_DM_periods_vs_p} occurs via a supercritical Hopf bifurcation, while the VL neurons are getting entrained via a subcritical bifurcation.
The study of codimension 2 points would be of interest in further studies of SCN networks.

This work was partially supported by the NNSF of China under Grant Nos. 10975053
and 11135001.

\end{document}